\documentclass[journal]{IEEEtran}

\usepackage[
    colorlinks=true,
    linkcolor=blue,
    citecolor=blue,
    urlcolor=blue,
    bookmarks
]{hyperref}
\usepackage{pgfplots}
\usepackage{multirow}
\usepackage{graphicx}
\usepackage{algorithm}
\usepackage{algpseudocode}
\usepackage{enumitem}
\usetikzlibrary{shapes.multipart,intersections}
\usepackage{cite}
\usepackage{amsmath,amssymb,amsfonts,amsthm,steinmetz}
\usepackage{mathrsfs}  
\usepackage{textcomp}
\usepackage{acronym}
\usepackage{xcolor}
\usepackage{upgreek,xspace}
\usepackage{array}
\usepackage{tikz}
\usetikzlibrary{calc}
\makeatletter
\newcommand{\gettikzxy}[3]{%
  \tikz@scan@one@point\pgfutil@firstofone#1\relax
  \edef#2{\the\pgf@x}%
  \edef#3{\the\pgf@y}%
}
\makeatother

\usepackage{comment}

\usepackage[draft]{todonotes}   
\usepackage[T1]{fontenc}

\usepackage{esvect}

\usepackage{times}
\usepackage{bm}
\usepackage{stmaryrd}
\usepackage{babel}
\usepackage{graphics, graphicx}
\usepackage{gensymb}
\usepackage{cite}
\usepackage{enumitem}
\usepackage{url}

\newcommand{\ma}[1]{\mathbf{#1}}
\newcommand{\bb}[1]{\mathbb{#1}}
\newcommand{\ten}[1]{\boldsymbol{\mathcal #1}}
\newcommand{\nmode}[2]{[#1]_{(#2)}}
\newcommand{\fronorm}[1]{\left\|#1\right\|_{\text{F}}}

\usepackage[all=normal,paragraphs=normal,floats=normal,mathspacing=normal,wordspacing=normal,charwidths=normal,mathdisplays=normal,leading=tight]{savetrees}

\begin{document}

\title{Channel Estimation via Tensor Decomposition for Dynamic Metasurface Antennas with Known Mutual Coupling: Algorithms and Experiments }

\author{Jean~Tapie,~\IEEEmembership{Graduate Student Member,~IEEE},~Bruno~Sokal,~\IEEEmembership{Member,~IEEE}, \\André~L.~F.~de~Almeida,~\IEEEmembership{Senior Member,~IEEE},~and~Philipp~del~Hougne,~\IEEEmembership{Member,~IEEE}
\thanks{
J.~Tapie and P.~del~Hougne are with Univ Rennes, CNRS, IETR - UMR 6164, F-35000, Rennes, France. P.~del~Hougne is also with the Department of Electronics and Nanoengineering, Aalto University, 00076 Espoo, Finland. (e-mail: jean.tapie@univ-rennes.fr; philipp.del-hougne@univ-rennes.fr)
}
\thanks{B.~Sokal and A.~L.~F.~de~Almeida are with the Wireless Telecom Research Group (GTEL), Department of Teleinformatics Engineering, Federal University of Ceara, 60416-200 Fortaleza-CE, Brazil (e-mail: brunosokal@gtel.ufc.br; andre@gtel.ufc.br).
}
\thanks{\textit{(Corresponding Author: Philipp del Hougne.)}}
\thanks{J.~Tapie and B.~Sokal contributed equally to this work.}
\thanks{This work was supported in part by the Nokia Foundation (project 20260028), the ANR France 2030 program (project ANR-22-PEFT-0005), the ANR PRCI program (project ANR-22-CE93-0010), the French Defense Innovation Agency (project 2024600), the European Union's European Regional Development Fund, the French region of Brittany and Rennes Métropole through the contrats de plan État-Région program (projects ``SOPHIE/STIC \& Ondes'' and ``CyMoCoD''), the National Institute of Science and Technology (INCT-Signals) sponsored by Brazil's National Council for Scientific and Technological Development (CNPq) (projects 406517/2022-3 and 303356/2025-1), and FUNCAP (project ITR-0214-00041.01.00/23).}
}

\maketitle

\begin{abstract}
Dynamic metasurface antennas (DMAs) are an emerging hybrid-MIMO technology distinguished by an ultrathin form factor, low cost, and low power consumption. In real-world DMA prototypes, mutual coupling (MC) between meta-elements is generally non-negligible; some architectures even deliberately exploit strong MC to enhance wave-domain flexibility. In this paper, we address channel estimation (CE) for DMAs with known MC by formulating it as a tensor-decomposition problem. We develop a generalized block Tucker alternating least squares (BTALS) algorithm, together with specialized variants for cases with known direct and/or feed channel. We also develop a reciprocity-aware bilinear factorization method for the case with known direct channel. We experimentally validate our algorithms using an 18~GHz DMA prototype whose 7 feeds and 96 meta-elements are strongly coupled via a chaotic cavity. Our general BTALS algorithm reaches an accuracy of 43.1~dB, only 0.3~dB below the upper bound imposed by experimental noise. All proposed algorithms generally outperform the prior-art reference scheme thanks to the superior noise rejection enabled by the tensor-based framework. We further study the minimum number of required measurements as a function of the number of feeds and demonstrate the importance of MC awareness by comparison with an MC-unaware benchmark. Finally, we apply BTALS to a second setup enabling the prediction of the DMA’s full dual-polarization 3D radiation diagram. We also measure the latter for DMA configurations optimized for channel-gain enhancement based on the estimated channels. Altogether, our work establishes the practical relevance of MC-aware tensor methods; beyond DMAs, it applies to all wireless systems with wave-domain programmability enabled by tunable lumped elements.
\end{abstract}


\section{Introduction}
\label{sec_introduction}
Hybrid antenna architectures with programmability in the wave domain are a technological enabler of massive multiple-input-multiple-output (mMIMO) wireless systems~\cite{del2026programmable}. To avoid the cost and energy consumption associated with requiring one radiofrequency (RF) chain per antenna element, these hybrid architectures operate with a significantly reduced number of RF chains, mapping the low-dimensional signal at the RF chains to a high-dimensional signal at the radiating elements via a programmable wave-domain transformation. Early proposals relied on reconfigurable analog combining boards connecting the reduced number of RF chains to a massive antenna array~\cite{zhang2005variable,venkateswaran2010analog,roh2014millimeter,han2015large,gong2020rf}. The recent emergence of dynamic metasurface antennas (DMAs) offers an alternative embodiment that is distinguished by its ultrathin form factor. A DMA is an ultrathin device that couples a low-dimensional number of feeds to a high-dimensional number of radiating meta-elements via microstrips or cavities~\cite{sleasman2015dynamic,yurduseven2018dynamically,yoo2018enhancing,sleasman2020implementation,del2020learned,boyarsky2021electronically,shlezinger2021dynamic,jabbar202460,yven2025end,prod2025beyond}. 
Maximizing the programmability of the wave-domain transformation implemented by a DMA is a natural goal in this context. Recent work has shown the benefits of strong inter-element coupling, which increases the sensitivity of the low-to-high-dimensional signal transformation to the configuration of tunable elements~\cite{prod2024mutual,MCbenefitsDMA}. In addition, beyond-diagonal DMAs were proposed to endow the inter-element coupling with programmability~\cite{prod2025beyond}.

To deploy a DMA in real-world conditions, wireless practitioners require an accurate system model that predicts signal transmission from feeds to users for any DMA configuration. Importantly, the system model must account for electromagnetic interactions among the tunable elements, especially in those DMA architectures that are intentionally designed to operate with strong inter-element interactions to boost wave-domain flexibility. An electromagnetically consistent system model relies on five groups of parameters~\cite{williams2022electromagnetic,tapie2025experimental}: 
\begin{enumerate}[label=(\roman*)]
    \item scattering characteristics of the possible states of the tunable elements,
    \item mutual coupling (MC) characteristics of the tunable elements,
    \item channels from the feeds to the tunable elements,
    \item channels from the tunable elements to the users,
    \item direct channels from the feeds to the users.
\end{enumerate}

It is often reasonable to assume that (i) and (ii) are known accurately by the wireless practitioner. If the deployed DMA prototype has been accurately simulated and there are no significant fabrication inaccuracies, (i) and (ii) are known~\cite{almunif2025network}. Alternatively, a set of operationally equivalent proxy parameters for (i) and (ii) can be estimated experimentally based on the fabricated prototype as described in~\cite{tapie2025experimental} (see Sec.~\ref{sec_SystemModel} for details on ambiguities in proxy parameters). 
It is important to recognize that an assumption of knowing (ii), or a proxy thereof, implies assuming that the MC between tunable elements does \textit{not} depend on the radio environment. In general, any environmental scattering object can contribute paths between pairs of tunable elements, thereby altering the MC between them. Indeed, in the context of wireless channels parametrized by reconfigurable intelligent surfaces (RISs), various experiments in the 2.45~GHz regime demonstrated strong variations in MC strength for the same RIS prototype in radio environments with different levels of environmental scattering~\cite{rabault2023tacit,del2025reducedrank}.
For DMA-based scenarios, however, environmental scattering objects are typically not in the immediate proximity of the DMA, and the path loss between the DMA's tunable elements and the environmental scattering object is typically high (especially at higher frequencies, such as 18~GHz and above). Consequently, the MC between the tunable elements is dominated by scattering within the DMA architecture, so it is approximately constant and independent of environmental scattering. In this case, the wireless practitioner can assume to know (ii).

During runtime, the wireless practitioner is thus left with the problem of MC-aware channel estimation (CE), i.e., estimating (iii), (iv), and (v) given knowledge of (i) and (ii). In addition, a two-time-scale MC-aware CE problem is relevant to DMAs because (iii) only needs to be estimated once in a given deployment scenario, whereas (iv) and (v) should be estimated once per coherence time. Thus, after an initial joint estimate of (iii), (iv), and (v), the problem of estimating (iv) and (v) given knowledge of (i), (ii), and (iii) is relevant.

Various theoretical papers have recently considered \textit{MC-unaware} CE problems for DMAs~\cite{zhang2023channel,yang2024extremely,10310000,10938032,magalhaes2025channel,11080457}, all under an assumption of negligible MC.\footnote{It is briefly mentioned in~\cite{10938032,11080457} that adding a fixed linear multiplication can account for MC; yet, this approach requires simplifications of the complete electromagnetically consistent system model (see details in Sec.~\ref{sec_SystemModel}).} Among these MC-unaware CE works for DMAs, \cite{10938032,magalhaes2025channel,11080457} leverage tensor decompositions, but for structurally different CE problems based on sparse geometric or separable propagation models, array-manifold assumptions, and/or OFDM-induced tensorizations, rather than for a full electromagnetically consistent multiport-network model that would be required for MC awareness.
To the best of our knowledge, the only work to date that has considered \textit{MC-aware} CE for DMAs is~\cite{tapie2025experimental}, where an algorithm that jointly estimates a set of operationally equivalent proxy parameters for (i), (ii), (iii), and (iv) is proposed and experimentally validated. Specifically, after a first phase in which the proxy parameters for (i), (ii), and (iii) are estimated, the technique in~\cite{tapie2025experimental} estimates a set of matching proxy parameters based on measurements of the end-to-end channel for one reference DMA configuration, single-element-flip DMA configurations, and random DMA configurations. The MC-aware CE technique for DMAs in~\cite{tapie2025experimental} does not fully exploit the tensor structure of the system model, and thus it cannot benefit from noise rejection enabled by tensor-decomposition techniques. 

Since the mathematical structure of an MC-aware model for a DMA-empowered system is the same as for an RIS-parametrized channel (see Sec.~\ref{sec_SystemModel}), MC-aware CE techniques for RISs are directly relevant to the present paper. Consequently, the same five aforementioned groups of parameters also apply to RISs.
On the one hand, multiple experimental works tackled the joint estimation of all five groups of parameters~\cite{sol2024experimentally,ContRIS_LWC,largeRIS_TCOM,del2025reducedrank}, motivated by scenarios in which fabrication inaccuracies or proprietary unknown RIS designs preclude perfect prior knowledge of any parameter. The same approach was also applied to a BD-RIS in~\cite{del2025physics}, noting that the same system model with the same five groups of parameters applies to a BD-RIS when a real-world implementation of the beyond-diagonal load circuit based on tunable lumped elements is considered. Moreover, the works in~\cite{sol2024experimentally,ContRIS_LWC,largeRIS_TCOM,del2025reducedrank} were the basis for the aforementioned MC-aware CE technique for DMAs~\cite{tapie2025experimental}.
On the other hand, several works on MC-aware CE for RISs assume prior knowledge of the RIS hardware, i.e., of the parameter sets (i) and (ii), and then exploit additional structural assumptions on the propagation model~\cite{Naffouri_MCaware,qiu2025mutual,JrCUP}. In particular, these works rely on geometry-based or sparse mmWave channel models, together with array-manifold knowledge and/or localization-oriented parameterizations. Their common focus is therefore on estimating a reduced set of physical channel parameters under a prescribed propagation structure, rather than recovering general unstructured channel matrices under the full multiport-network model. The underlying assumptions may be restrictive in practice, since accurate array-manifold knowledge can be compromised by antenna-position uncertainties, calibration errors, RF phase drifts, near-field propagation, and other model-reality mismatches.
Among these MC-aware CE works, only \cite{JrCUP} leverages tensor methods, but for a purpose that fundamentally differs from our goal in this work: the tensor-ESPRIT stage in \cite{JrCUP} exploits rotational-invariance in a geometry-based localization model to estimate physical signal parameters. In contrast, the focus of the present work is on the estimation of general channel matrices under the multiport-network model, i.e., of the parameter sets (iii), (iv), and (v), without relying on array-manifold or localization-specific assumptions.

To the best of our knowledge, no prior work has exploited tensor decomposition for MC-aware CE without structural assumptions on the channels, neither for DMAs nor RISs. This gap is important because of its practical relevance and the potential of tensor methods to naturally leverage the multidimensional data structures arising from the measurements collected across multiple DMA configurations, feeds, and users. Tensor decompositions have a long track record in wireless communications~\cite{deAlmeida2007,FavierAlmeida2014JASP,deAlmeida2008confac,deAlmeidaFavier2013,FavierAlmeida2014TSP,Ximenes_2015} and recent theoretical works have shown strong potential for programmable wave-domain systems such as RISs, BD-RISs, and DMAs~\cite{Araujo2021JSTSP,Araujo2020SAM,bdris_ce_tensor,Sokal2024,10938032,magalhaes2025channel,11080457}. 
While these theoretical results highlight the potential of tensor-based methods for addressing complex CE problems in programmable wireless environments, they all make idealized assumptions by neglecting real-world hardware constraints, especially the MC phenomena that inevitably arise in real-world prototypes.

In this paper, we fill the gap on MC-aware CE via tensor decomposition for DMAs based on a multiport-network model without assumptions on the structure of the involved channels. Our work combines electromagnetically consistent system modeling and modern multidimensional signal processing techniques, and connects them with extensive experimental validation.
Our contributions are summarized as follows.
\begin{enumerate}
    \item We introduce a tensor-based formulation of MC-aware CE for DMAs with known MC, based on an electromagnetically consistent multiport-network model and covering four practical levels of prior channel knowledge.
    
    \item We propose a block Tucker alternating least squares (BTALS) algorithm for joint recovery of the unknown channels, as well as specialized BTALS variants that exploit prior knowledge of the direct channel and/or the feed channel.
    
    \item We propose a reciprocity-aware bilinear factorization method for the known-direct-channel case.
    
    \item We experimentally validate, for the first time, tensor-decomposition-based MC-aware CE on a real-world programmable wave-domain system, using an 18~GHz strongly coupled DMA prototype in two distinct setups.
    
    \item We systematically study the dependence of our algorithms' CE performance on the number of available measurements and the number of DMA feeds. We further benchmark our algorithms against the MC-unaware case.
    
    \item We demonstrate the usefulness of the estimated channel parameters in model-based optimization for channel gain maximization, with experimental measurements confirming the predicted performance.
\end{enumerate}

\textit{Organization:} 
In Sec.~\ref{sec_SystemModel}, we describe the DMA-empowered system model, including a discussion of parameter ambiguities. 
In Sec.~\ref{sec_ProblemStatement}, we describe the four considered problem statements on MC-aware CE for DMAs.
In Sec.~\ref{sec_tensor_formulation}, we introduce our tensor-based system formulation.
In Sec.~\ref{sec_CEalgorithms}, we present our algorithms for MC-aware CE via tensor decomposition.
In Sec.~\ref{sec_ExpVal}, we experimentally validate our algorithms. Specifically, we describe our DMA prototype, experimental setups, measurement procedure, experimentally verifiable CE accuracy metrics, CE results, and performance evaluation in DMA optimization.
We close with a brief discussion in Sec.~\ref{sec_Discussion} and conclusion in Sec.~\ref{sec_Conclusion}.

\textit{Notation:}

$\mathbb{C}$ and $\mathbb{B}\triangleq\{0,1\}$ denote the sets of complex and binary numbers, respectively.
$(\cdot)^\top$, $(\cdot)^H$, and $(\cdot)^+$ denote transpose, Hermitian transpose, and Moore--Penrose pseudoinverse, respectively.
$|\cdot|$, $\|\cdot\|_2$, and $\fronorm{\cdot}$ denote the absolute value, Euclidean norm, and Frobenius norm, respectively.
$\mathbf{I}_a$ denotes the $a\times a$ identity matrix.
$\mathbf 0$ and $\mathbf 1$ denote all-zeros and all-ones vectors/matrices of appropriate sizes.
$\mathrm{diag}(\mathbf{a})$ denotes the diagonal matrix whose diagonal entries are given by the vector $\mathbf{a}$, and $\mathrm{blkdiag}(\cdot)$ denotes block-diagonal concatenation.
$\otimes$ denotes the Kronecker product.
$\mathrm{vec}(\cdot)$ and $\mathrm{vech}(\cdot)$ denote vectorization and half-vectorization, respectively.
$A_{i,j}$ denotes the $(i,j)$th entry in the matrix $\mathbf{A}$.
$\mathbf{A}_{\mathcal{B}\mathcal{C}}$ denotes the submatrix (block) of $\mathbf{A}$ selected by the row indices $\mathcal{B}$ and the column indices $\mathcal{C}$.

\textit{Tensor Algebra Preliminaries:} 
Let $\ma{T}_k \in \bb{C}^{I \times J}$, $k=1,\ldots,K$, and define the third-order tensor $\ten{T} = [\ma{T}_1 \sqcup_3 \cdots \sqcup_3 \ma{T}_K] \in \bb{C}^{I \times J \times K}$, where $\sqcup_3$ denotes concatenation along the third mode. Its mode-$1$, mode-$2$, and mode-$3$ unfoldings are
\begin{align*}
\nmode{\ten{T}}{1} &= [\ma{T}_1,\ldots,\ma{T}_K] \in \bb{C}^{I \times JK},\\
\nmode{\ten{T}}{2} &= [\ma{T}_1^{\top},\ldots,\ma{T}_K^{\top}] \in \bb{C}^{J \times IK},\\
\nmode{\ten{T}}{3} &= [\mathrm{vec}(\ma{T}_1),\ldots,\mathrm{vec}(\ma{T}_K)]^{\top} \in \bb{C}^{K \times IJ}.
\end{align*}
Given $\ma{A} \in \bb{C}^{L \times I}$, $\ma{B} \in \bb{C}^{M \times J}$, and $\ma{C} \in \bb{C}^{N \times K}$, the multilinear transform
\[
\ten{Y} = \ten{T} \times_1 \ma{A} \times_2 \ma{B} \times_3 \ma{C} \in \bb{C}^{L \times M \times N},
\]
where $\times_n$ denotes the mode-$n$ product, satisfies
\begin{align*}
\nmode{\ten{Y}}{1} &= \ma{A}\,\nmode{\ten{T}}{1}(\ma{C}\otimes\ma{B})^{\top},\\
\nmode{\ten{Y}}{2} &= \ma{B}\,\nmode{\ten{T}}{2}(\ma{C}\otimes\ma{A})^{\top},\\
\nmode{\ten{Y}}{3} &= \ma{C}\,\nmode{\ten{T}}{3}(\ma{B}\otimes\ma{A})^{\top}.
\end{align*}
More extensive background on tensor algebra notations, products, and decompositions can be found in~\cite{kolda,sidiropoulos2017tensor}.

\section{System Model}
\label{sec_SystemModel}

\begin{figure}
    \centering
    \includegraphics[width=\columnwidth]{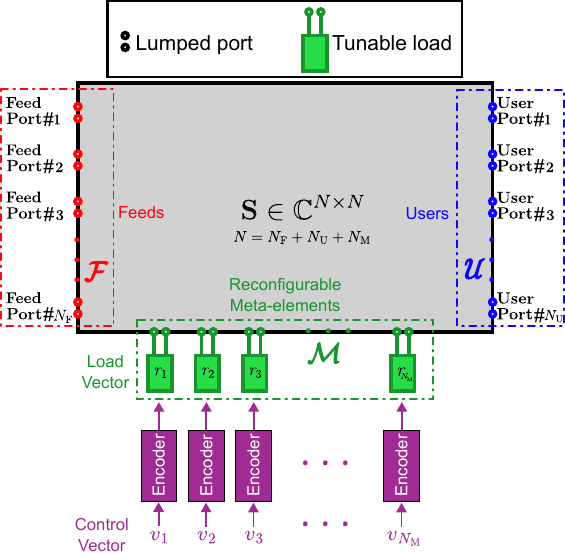}
    \caption{MNT system model for a DMA with $N_\mathrm{F}$ feeds and $N_\mathrm{M}$ reconfigurable meta-elements serving $N_\mathrm{U}$ users. The encoding of the control vector $\mathbf{v}$ in the load vector $\mathbf{r}$ that enters the MNT formalism is highlighted.}
    \label{Fig1}
\end{figure}

In this section, we describe how our system model maps the binary control vector determining the configuration of the DMA's $N_\mathrm{M}$ 1-bit-programmable meta-elements to the system's corresponding $N_\mathrm{U}\times N_\mathrm{F}$ end-to-end multiple-input-multiple-output (MIMO) transfer function from the $N_\mathrm{F}$ DMA feeds to the $N_\mathrm{U}$ ports of the users. We begin by explaining how the binary control vector is mapped to the load vector, which contains the physical scattering properties of the DMA's meta-elements. Next, we use MNT to map the load vector to the end-to-end channel matrix. Then, we discuss the inevitable ambiguities that arise when all model parameters are estimated experimentally, as well as the  operational irrelevance of these ambiguities.

Our starting point is a partition of the system into three entities: (i) ports via which waves are injected and/or received; (ii) tunable lumped elements; (iii) static scattering objects. The ports in (i) comprise the ports of the DMA's feeds and the ports of the users' antennas. Each tunable lumped element in (ii) can be described as a ``virtual'' additional port terminated by an individual tunable load. The static scattering objects in (iii) comprise the scattering structures inside the DMA and the structural scattering of the radiating meta-elements, as well as, if applicable, any environmental scattering and structural scattering of the user antennas. 

\textit{Encoding Function:} The control vector  $\mathbf{v}\in\mathbb{B}^{N_\mathrm{M}}$ defines the configuration of the $N_\mathrm{M}$ 1-bit-programmable lumped elements. The load vector $\mathbf{r}\in\mathbb{C}^{N_\mathrm{M}}$ contains the reflection coefficients of the $N_\mathrm{M}$ individual loads associated with the description of the tunable lumped elements as ``virtual'' ports terminated by individual loads. In line with our prototype, we assume that all $N_\mathrm{M}$ tunable lumped elements are identical and independently tunable with two possible states. The two states are characterized by the reflection coefficients $\alpha\in\mathbb{C}$ and $\beta\in\mathbb{C}$ of the associated loads. In this case, the mapping $\mathbf{v} \rightarrow \mathbf{r}$ is affine~\cite{tapie2025experimental}:
\begin{equation}
    \mathbf{r}(\mathbf{v}) = \alpha \,\mathbf{1} + (\beta-\alpha)\, \mathbf{v}.
    \label{eq_encod}
\end{equation}

\textit{MNT Function:}
The ensemble of the $N_\mathrm{M}$ tunable loads constitutes an $N_\mathrm{M}$-port system which is fully characterized by its scattering matrix $\mathbf{\Phi}(\mathbf{r})=\mathrm{diag}(\mathbf{r})$. Moreover, the ensemble of all static scattering objects constitutes an $N$-port system, where $N=N_\mathrm{F}+N_\mathrm{M}+N_\mathrm{U}$, which is fully characterized by its scattering matrix $\mathbf{S}\in\mathbb{C}^{N\times N}$. We use the same reference impedance of $Z_0=50\ \Omega$ at all ports to define all scattering parameters in this paper. Moreover, we assume that signal generators and detectors are matched to $Z_0$. The $N_\mathrm{M}$-port system terminates the $N_\mathrm{M}$ ``virtual'' ports of the $N$-port system. The end-to-end channel matrix $\mathbf{H}\in\mathbb{C}^{N_\mathrm{U}\times N_\mathrm{F}}$ from feeds to user ports resulting from this connection is given by standard MNT~\cite{tapie2025experimental}: 
\begin{equation}
\mathbf{H}(\mathbf{r}) = \mathbf{H}_0 + \mathbf{A}\,\bigl(\mathbf{I}_{N_\mathrm{M}}\,-\,\mathbf{\Phi}(\mathbf{r})\,\mathbf{\Gamma}\bigr)^{-1}\,\mathbf{\Phi}(\mathbf{r})\,\mathbf{B},
\label{eq_MNT}
\end{equation}
where, for notational ease, $\mathbf{H}_0 \triangleq \mathbf{S}_\mathcal{UF}\in\mathbb{C}^{N_\mathrm{U}\times N_\mathrm{F}}$, $\mathbf{A} \triangleq \mathbf{S}_\mathcal{UM}\in\mathbb{C}^{N_\mathrm{U}\times N_\mathrm{M}}$, $\mathbf{\Gamma} \triangleq \mathbf{S}_\mathcal{MM}\in\mathbb{C}^{N_\mathrm{M}\times N_\mathrm{M}}$, and $\mathbf{B} \triangleq \mathbf{S}_\mathcal{MF}\in\mathbb{C}^{N_\mathrm{M}\times N_\mathrm{F}}$, and  $\mathcal{F}$, $\mathcal{U}$, and $\mathcal{M}$ denote the sets containing the port indices associated with the feed ports, the user ports, and ``virtual'' ports, respectively.
Typical DMAs are built exclusively from reciprocal components and are therefore reciprocal. Consequently, reciprocity imposes $\mathbf{\Gamma} = \mathbf{\Gamma}^\top$. 

\textit{Benchmark MC-Unaware Model:} To examine the importance of MC awareness, we also consider a simple MC-unaware model that can be obtained from (\ref{eq_MNT}) by assuming $\mathbf{\Gamma}=\mathbf{0}$:
\begin{equation}
\mathbf{H}^\mathrm{noMC}(\mathbf{r}) = \mathbf{H}_0 + \mathbf{A}\,\mathbf{\Phi}(\mathbf{r})\,\mathbf{B}.
\label{eq_MC_unaware}
\end{equation}

\textit{Parameter Ambiguities:}
As mentioned in the introduction, our system model involves five groups of parameters: (i) the scattering characteristics of the tunable loads' possible states (i.e., $\alpha$ and $\beta$), (ii) the MC characteristics of the tunable elements (i.e., $\mathbf{\Gamma}$), (iii) the channels from feeds to ``virtual'' ports (i.e., $\mathbf{B}$), (iv) the channels from ``virtual'' ports to user ports (i.e., $\mathbf{A}$), and (v) the direct channels from feeds to user ports (i.e., $\mathbf{H}_0$). If all system details are perfectly known (DMA architecture, radio environment, user locations and orientations), then all MNT parameters can be obtained with a single full-wave simulation~\cite{tapie2023systematic,zheng2024mutual,almunif2025network} (unless the simulation is computationally too costly). Due to fabrication inaccuracies, unknown details of the radio environment, and/or unknown user locations, these conditions are not met in real-world settings, so the model parameters need to be estimated experimentally. However, one cannot simply measure the model parameters because the ``virtual'' ports are not connectorized and too numerous. Indirect estimations of the model parameters can identify the MNT parameters unambiguously only if the encoding function's parameters are known and the encoding function satisfies specific conditions (at least three distinct terminations per ``virtual'' port, as well as coupled terminations between neighboring ``virtual'' ports and at least one ``virtual'' and one accessible port~\cite{del2024virtual2p0,del2025virtual3p0,del2025wireless}). Our system model's encoding function does not satisfy these criteria, and we do not know $\alpha$ or $\beta$. Consequently, we can only estimate a set of proxy model parameters that are subject to various ambiguities (diagonal similarity, complex scaling, M\"obius transformation -- see [Sec.~III,~\cite{salmi2026electromagnetically}] for details). However, since these ambiguities do not impact the mapping $\mathbf{v}\rightarrow \mathbf{H}$, they are operationally irrelevant for model-based optimization of the DMA configuration to impose desired properties on the end-to-end channel matrix (e.g., beamforming). We denote proxy parameters with a tilde in this paper to distinguish them from ground-truth parameters.

\section{Problem Statement}
\label{sec_ProblemStatement}

In this paper, we consider four types of MC-aware CE problems for DMAs; an overview is provided in Table~\ref{tab:ce_types}. In all cases, we assume that a matching set of proxy parameters for the encoding function (i.e., $\tilde{\alpha}$ and $\tilde{\beta}$) and the inter-element MC matrix (i.e., $\tilde{\mathbf{\Gamma}}$) is known from a one-off experimental characterization of the DMA prototype with the procedure in~\cite{tapie2025experimental}. This characterization could have been conducted in the current setup or a different setup, as long as the same DMA prototype was used, because we assume that $\tilde{\alpha}$, $\tilde{\beta}$, and $\tilde{\mathbf{\Gamma}}$ do not depend on the radio environment.
As explained in the introduction, this assumption is reasonable due to significant path loss at our operating frequency of 18~GHz and the absence of environmental scatterers in close proximity to the DMA.
Importantly, the ambiguities in $\tilde{\alpha}$, $\tilde{\beta}$, and $\tilde{\mathbf{\Gamma}}$ must harmonize, which they naturally do if the proxy parameters are jointly obtained with the procedure described in~\cite{tapie2025experimental}.

\begin{table}[b]
\centering
\caption{Overview of the four considered MC-aware CE problem types.}
\label{tab:ce_types}
\renewcommand{\arraystretch}{1.15}
\setlength{\tabcolsep}{6pt}
\begin{tabular}{lccccc}
\hline
\textbf{Type} 
& $\tilde{\alpha},\tilde{\beta},\tilde{\mathbf{\Gamma}}$ 
& $\tilde{\mathbf{H}}_0$ 
& $\tilde{\mathbf{A}}$ 
& $\tilde{\mathbf{B}}$ \\
\hline
Type 1 & known & TBD & TBD & TBD \\
Type 2 & known & known & TBD & TBD \\
Type 3 & known & TBD & TBD & known \\
Type 4 & known & known & TBD & known \\
\hline
\end{tabular}
\end{table}

For \textit{Type 1}, we assume that $\tilde{\alpha}$, $\tilde{\beta}$, and $\tilde{\mathbf{\Gamma}}$ are known, and we jointly estimate $\tilde{\mathbf{H}}_0$, $\tilde{\mathbf{A}}$, and $\tilde{\mathbf{B}}$. This is the general CE setting.

The only difference for \textit{Type 2} with respect to \textit{Type 1} is that we assume to additionally know $\tilde{\mathbf{H}}_0$. This situation is relevant because with the choice $\tilde{\alpha}=0$ made in~\cite{tapie2025experimental} we can obtain an estimate of $\tilde{\mathbf{H}}_0$ by measuring the end-to-end channel matrix for a single reference DMA configuration: $\tilde{\mathbf{H}}_0 = \mathbf{H}(\mathbf{v}_0)$, where $\mathbf{v}_0=\mathbf{0}$.\footnote{If $\tilde{\alpha}\neq 0$, one can apply an ambiguity transformation of the proxy parameters to another set of proxy parameters for which $\tilde{\alpha}=0$. An example of a related ambiguity transformation is available in [(21),~\cite{salmi2026electromagnetically}].}
The relevant CE problem is thus to estimate $\tilde{\mathbf{A}}$ and $\tilde{\mathbf{B}}$ given $\tilde{\alpha}$, $\tilde{\beta}$, $\tilde{\mathbf{\Gamma}}$, and $\tilde{\mathbf{H}}_0$.

With \textit{Type 3} and \textit{Type 4}, we capture the natural two-time-scale structure of DMA-empowered links. For a fixed deployment of the DMA hardware, the channels from feeds to meta-elements are typically quasi-static over long time scales, whereas the channels from meta-elements to users and from feeds to users vary with the users' locations and orientations and with the radio environment's dynamics. Accordingly, after the procedure described in~\cite{tapie2025experimental}, or after running a \textit{Type 1} or \textit{Type 2} procedure once, we can treat $\tilde{\mathbf{B}}$ as known. If $\tilde{\mathbf{H}}_0$ is not assumed known, the resulting per-coherence CE problem corresponds to \textit{Type 3}, in which $\tilde{\mathbf{H}}_0$ and $\tilde{\mathbf{A}}$ are estimated. If, in addition, $\tilde{\mathbf{H}}_0$ is obtained from a single reference-configuration measurement as in \textit{Type 2}, the problem of \textit{Type 3} reduces to \textit{Type 4}, in which only $\tilde{\mathbf{A}}$ remains to be estimated.

\textbf{Remark:}
Although $\tilde{\mathbf{H}}_0$ can be estimated with a single reference-configuration measurement as in \textit{Type 2} and in \textit{Type 4}, we do not assume it known in \textit{Type 1} and \textit{Type 3} in order to systematically assess the trade-off between estimation accuracy and computational complexity in the comparison between \textit{Type 1} and \textit{Type 2} as well as in the comparison between \textit{Type 3} and \textit{Type 4}.

\section{Tensor-Based System Model Formulation}
\label{sec_tensor_formulation}

In this section, we present a tensor formulation based on \eqref{eq_MNT} that we will exploit in Sec.~\ref{sec_CEalgorithms} to derive our proposed algorithms for CE.

Our CE techniques are based on $K$ measurements of the end-to-end channel matrix, each using a different DMA configuration. For the $k$th DMA configuration, we denote the control vector by $\mathbf{v}_k$. The corresponding load vector is $\mathbf{r}_k \triangleq \mathbf{r}(\mathbf{v}_k)$ and the corresponding scattering matrix of the ensemble of tunable loads is $\mathbf{\Phi}_k \triangleq \mathbf{\Phi}(\mathbf{r}_k)=\mathrm{diag}(\mathbf{r}_k)$. The corresponding end-to-end channel matrix is
\begin{equation}
\label{eq:Hk_factorized}
\mathbf{H}_k \triangleq \mathbf{H}(\mathbf{r}_k)
= \mathbf{H}_0 + \mathbf{A}\,\mathbf{\Omega}_k\,\mathbf{B},
\end{equation}
where
\begin{equation}
\label{eq:Sk_def}
\mathbf{\Omega}_k \triangleq \bigl(\mathbf{I}_{N_\mathrm{M}}-\mathbf{\Phi}_k\,\mathbf{\Gamma}\bigr)^{-1}\mathbf{\Phi}_k
\in\mathbb{C}^{N_\mathrm{M}\times N_\mathrm{M}}.
\end{equation}
As mentioned, typical DMAs are reciprocal, which implies $\mathbf{\Gamma} = \mathbf{\Gamma}^\top$ and thus  $\mathbf{\Omega}_k = \mathbf{\Omega}_k^\top$. 

When neglecting $\ma{H}_0$, one can interpret the mathematical structure of \eqref{eq:Hk_factorized} as a frontal slice of a Tucker-$2$ tensor, as shown in \cite{bdris_ce_tensor}. In the general case with non-negligible $\ma{H}_0$, we can recast \eqref{eq:Hk_factorized} as follows:
\begin{align}\label{eq:Hk_factorized_slice_tucker}
   \mathbf{H}_k  
&= \left[\mathbf{H}_0,\, \ma{A}\right]\begin{bmatrix} \ma{I}_{N_{\text{F}}} &  \\ 
 & \ma{\Omega}_k\end{bmatrix}  \begin{bmatrix}
     \ma{I}_{N_{\text{F}}} \\
     \mathbf{B}
 \end{bmatrix}  \\
 &= \ma{C} \, \ma{\bar{\Omega}}_{k}\, \ma{D},\label{eq:Hk_factorized_slice_tucker_compact}
\end{align}
where $\ma{C} = \left[\mathbf{H}_0,\, \ma{A}\right] \in \mathbb{C}^{N_{\text{U}} \times (N_{\text{F}} + N_{\text{M}})}$,  $\ma{\bar{\Omega}}_{k} = \text{blkdiag}(\ma{I}_{N_{\text{F}}},\ma{\Omega}_k) \in \mathbb{C}^{ (N_{\text{F}} + N_{\text{M}}) \times  (N_{\text{F}} + N_{\text{M}})}$, and $\ma{D} =[  \ma{I}_{N_{\text{F}}}, \mathbf{B}]^\top \in \mathbb{C}^{ (N_{\text{F}} + N_{\text{M}}) \times  N_{\text{F}} }$. Therefore, by concatenating the equivalent channel $\ma{H}_k$ over all $K$ measurements, across the third dimension,  the equivalent channel tensor is given as: 
\begin{align}
    \ten{H} &= \bar{\ten{T}} \times_1 \ma{C} \times_2 \ma{D}^{\top} \times_3 \ma{I}_K \in \bb{C}^{N_{\text{U}}  \times N_{\text{F}} \times K },
\end{align}
where the equivalent channel tensor $\ten{H}$ concatenates  $\mathbf{H}_k $ across the third dimension, i.e., $\ten{H} = [\mathbf{H}_1  \sqcup_3 \ldots \sqcup_3\mathbf{H}_K ]$. Likewise,  $\bar{\ten{T}}$ is formed by concatenating the DMA factor $\ma{\bar{\Omega}}_{k}$, across the third dimension, i.e., $\bar{\ten{T}} = [\ma{\bar{\Omega}}_{1} \sqcup_3 \ldots \sqcup_3 \ma{\bar{\Omega}}_{K}] \in \mathbb{C}^{ (N_{\text{F}} + N_{\text{M}}) \times  (N_{\text{F}} + N_{\text{M}}) \times K}$. The three unfoldings of ${\ten{H}}$ are 
\begin{align}
 \label{eq:tenH_1}   \nmode{\ten{H}}{1} &= \ma{C}\,\nmode{\bar{\ten{T}}}{1}\big(\ma{I}_K \otimes \ma{D}^{\top}\big)^\top \in \bb{C}^{N_{\text{U}} \times N_{\text{F}}  K  }, \\
    \label{eq:tenH_2}\nmode{\ten{H}}{2} &= \ma{D}^{\top}\,\nmode{\bar{\ten{T}}}{2}\big(\ma{I}_K \otimes \ma{C}\big)^\top \in \bb{C}^{N_{\text{F}} \times N_{\text{U}}  K  }, \\
    \label{eq:tenH_3}\nmode{\ten{H}}{3} &= \nmode{\bar{\ten{T}}}{3}\big(\ma{D}^{\top} \otimes \ma{C}\big)^\top \in \bb{C}^{K \times N_{\text{U}}N_{\text{F}}    }. 
\end{align}

\section{Algorithms for MC-Aware CE via \\ Tensor Decomposition}
\label{sec_CEalgorithms}

In this section, we present two main algorithms and their derivations for estimating the involved channels across the four MC-aware problem types, as described in Table~\ref{tab:ce_types}.
To simplify the notation, we write this section in terms of the true physical parameters rather than their proxy counterparts, denoted by tildes. In the experimental setting considered in this paper, however, the known parameters are in general ambiguous proxy parameters, as discussed in Sec.~\ref{sec_SystemModel} and Sec.~\ref{sec_ExpVal}.

\subsection{Type 1 MC-Aware CE Problem}
\label{subsec_Type1}
\textit{BTALS-I:} Based on \cite{bdris_ce_tensor}, we develop a generalized BTALS algorithm for cases with non-negligible direct channel and non-negligible MC. Our BTALS algorithm iteratively solves two least-squares (LS) problems that exploit the unfoldings \eqref{eq:tenH_1} and \eqref{eq:tenH_2}:
\begin{align}
    \label{eq:als_prob_C}\hat{\ma{C}} &=  \underset{\ma{C}}{\text{argmin}}\fronorm{\nmode{{\ten{H}}}{1} - \ma{C}\,\nmode{\bar{\ten{T}}}{1}(\ma{I}_K \otimes \ma{D}^{\top})^\top }^2, \\
    \label{eq:als_prob_D}\hat{\ma{D}}^{\top} &=  \underset{\ma{D}^{\top}}{\text{argmin}}\fronorm{\nmode{{\ten{H}}}{2} - \ma{D}^{\top}\,\nmode{\bar{\ten{T}}}{2}(\ma{I}_K \otimes \ma{C})^\top }^2.
\end{align}
The corresponding solutions are
\begin{align}
    \label{eq:btals_sol_C} \hat{\ma{C}} &= \nmode{{\ten{H}}}{1}\big(\nmode{\bar{\ten{T}}}{1}(\ma{I}_K \otimes \ma{D}^{\top})^\top\big)^+, \\
    \label{eq:btals_sol_D} \hat{\ma{D}}^{\top} &= \nmode{{\ten{H}}}{2}\big(\nmode{\bar{\ten{T}}}{2}(\ma{I}_K \otimes \ma{C})^\top\big)^+.
\end{align}
A necessary dimensional condition for the uniqueness of the LS solution in \eqref{eq:btals_sol_C} and \eqref{eq:btals_sol_D} is  $K \geq \max\{1 + \frac{N_{\text{M}}}{N_{\text{F}}},\,\,  \frac{N_{\text{F}}+N_{\text{M}}}{N_{\text{U}}} \}$.

    \begin{algorithm}[b]
\caption{Generalized BTALS Algorithm for \textit{Type 1}}
\label{alg:BTALS}
\begin{algorithmic}[1]
    \State \textbf{Input:} $\ten{H}$, $\bar{\ten{T}} $, $\hat{\ma{B}}^{0}$, max. number of iterations $I$. Set $i=1$.
    \For{$i = 1,\ldots,I$}
        \State Compute an estimate for $\hat{\ma{C}}^{i}$ using \eqref{eq:btals_sol_C}.
        \State Compute an estimate for $\left({\hat{\ma{D}}^i}\right)^\top$ using \eqref{eq:btals_sol_D}.
        \State Replace upper $N_{\text{F}}\times N_{\text{F}}$ block of $\hat{\ma{D}}^{i}$ with $\ma{I}_{N_{\text{F}}}$.
    \EndFor
    \State \textbf{return} $\hat{\ma{C}} = \hat{\ma{C}}^{i}  $, $\hat{\ma{D}} = \hat{\ma{D}}^{i}  $. 
\end{algorithmic}
\end{algorithm}

Our BTALS algorithm is summarized in Algorithm~\ref{alg:BTALS}.
In the first iteration $i=1$, the feed channel can be randomly initialized as $\hat{\ma{B}}^{0}$ to form $\hat{\ma{D}}^0 =[ \ma{I}_{N_{\text{F}}}, \mathbf{\hat{B}}^{0}]^\top$, which is substituted into \eqref{eq:als_prob_C} to obtain $\hat{\ma{C}}^{1}$. Then, $\hat{\ma{C}}^{1}$ is substituted into  \eqref{eq:als_prob_D} to obtain $\hat{\ma{D}}^{1}$. This process is repeated until a maximum number of iterations or a convergence threshold is reached. 
Since the LS update in \eqref{eq:btals_sol_D} is unconstrained, the estimate $\hat{\ma{D}}^{i}$ may be obtained as a generic full matrix. However, the true factor $\ma{D}$ has the structured form $\ma{D} =[  \ma{I}_{N_{\text{F}}}, \mathbf{B}]^\top$. 
Therefore, after each iteration, we enforce this structure by replacing the upper $N_\mathrm{F}\times N_\mathrm{F}$ block of $\hat{\ma{D}}^{i}$ with $\ma{I}_{N_\mathrm{F}}$, yielding $\hat{\ma{D}}^{i} =[  \ma{I}_{N_{\text{F}}}, \hat{\ma{B}}^{i}]^\top$.

The factorization in Algorithm~\ref{alg:BTALS} is not unique because only the product $\mathbf{A}\,\mathbf{\Omega}_k\,\mathbf{B}$ is identified. For any nonzero scalar $\gamma\in\mathbb{C}$, replacing $\mathbf{A}$ by $\gamma\mathbf{A}$ and $\mathbf{B}$ by $\gamma^{-1}\mathbf{B}$ leaves $\mathbf{A}\,\mathbf{\Omega}_k\,\mathbf{B}$ unchanged for every $k$. Since $\mathbf{D}=[\mathbf{I}_{N_\mathrm{F}},\,\mathbf{B}]^\top$ contains a fixed identity block, this ambiguity does not affect $\mathbf{H}_0$. Hence, $\mathbf{H}_0$ is identified directly, whereas $\mathbf{A}$ and $\mathbf{B}$ are identified only up to a reciprocal scalar ambiguity: $\hat{\ma{H}}_0 \approx \ma{H}_0,\ \ma{A} \approx \gamma\hat{\ma{A}}$, and $\ma{B} \approx \gamma^{-1}\hat{\ma{B}}$. This ambiguity is operationally irrelevant because it does not affect the mapping from the control vector to the end-to-end channel matrix.

\subsection{Type 2 MC-Aware CE Problem}
\label{subsec_Type2}

With known $\mathbf{H}_0$, we can define
\begin{align}
  \label{eq:Hk_factorized_type2}  \mathring{\mathbf{H}}_k \triangleq \mathbf{H} (\mathbf{r}_k) - \mathbf{H}_0 = \mathbf{A}\,\mathbf{\Omega}_k\,\mathbf{B}.
\end{align}
Analogous to Sec.~\ref{sec_tensor_formulation}, the channel tensor structure is given as
\begin{align}
     \label{eq:Hk_factorized_type2_ten} \mathring{\ten{H}} = \ten{T} \times_1 \ma{A} \times_2 \ma{B}^{\top} \times_3 \ma{I}_{K} \in  \bb{C}^{N_{\text{U}}  \times N_{\text{F}} \times K }, 
\end{align}
where $\mathring{\ten{H}} = [\mathring{\mathbf{H}}_1  \sqcup_3 \ldots \sqcup_3\mathring{\mathbf{H}}_K ]$ and $\ten{T} = [\ma{{\Omega}}_{1} \sqcup_3 \ldots \sqcup_3 \ma{{\Omega}}_{K}] \in \mathbb{C}^{  N_{\text{M}} \times   N_{\text{M}} \times K}$. 
The three unfoldings of $\mathring{\ten{H}}$ are
\begin{align}
  \label{eq:type_2_unf_1}  \nmode{\mathring{\ten{H}}}{1} &= \ma{A}\,\nmode{\ten{T}}{1}(\ma{I}_{K} \otimes \ma{B}^{\top})^\top \in \bb{C}^{N_{\text{U}}  \times N_{\text{F}} K}, \\
    \label{eq:type_2_unf_2}\nmode{\mathring{\ten{H}}}{2} &= \ma{B}^{\top}\,\nmode{\ten{T}}{2}(\ma{I}_{K} \otimes \ma{A})^\top \in \bb{C}^{N_{\text{F}}  \times N_{\text{U}} K}, \\
    \label{eq:type_2_unf_3}\nmode{\mathring{\ten{H}}}{3} &= \nmode{\ten{T}}{3}(\ma{B}^{\top} \otimes \ma{A})^\top \in \bb{C}^{K \times N_{\text{U}}   N_{\text{F}}}. 
\end{align}
We now propose two methods to solve the \textit{Type 2} MC-aware CE problem by exploiting this tensor structure. 

\textit{BTALS-II:}
We specialize BTALS-I to the case of a known $\mathbf{H}_0$.
Analogous to Sec.~\ref{subsec_Type1}, 
we solve two LS problems, this time using the $\nmode{\mathring{\ten{H}}}{1}$ and $\nmode{\mathring{\ten{H}}}{2}$:
\begin{align}
    \label{eq:als_prob_A}\hat{\ma{A}} &=  \underset{\ma{A}}{\text{argmin}}\fronorm{\nmode{{\mathring{\ten{H}}}}{1} - \ma{A}\,\nmode{{\ten{T}}}{1}(\ma{I}_K \otimes \ma{B}^{\top})^\top }^2, \\
    \label{eq:als_prob_B}\hat{\ma{B}}^{\top} &=  \underset{\ma{B}^{\top}}{\text{argmin}}\fronorm{\nmode{{\mathring{\ten{H}}}}{2} - \ma{B}^{\top}\,\nmode{{\ten{T}}}{2}(\ma{I}_K \otimes \ma{A})^\top }^2.
\end{align}
The corresponding solutions are 
\begin{align}
    \label{eq:btals_sol_A} \hat{\ma{A}} &= \nmode{{\mathring{\ten{H}}}}{1}\big(\nmode{{\ten{T}}}{1}(\ma{I}_K \otimes \ma{B}^{\top})^\top\big)^+, \\
    \label{eq:btals_sol_B} \hat{\ma{B}}^{\top} &= \nmode{{\mathring{\ten{H}}}}{2}\big(\nmode{{\ten{T}}}{2}(\ma{I}_K \otimes \ma{A})^\top\big)^+.
\end{align}
A necessary dimensional condition for the uniqueness of the LS solution in \eqref{eq:btals_sol_A} and  \eqref{eq:btals_sol_B} is $K \geq \max\{ \frac{N_{\text{M}}}{N_{\text{U}}},\,\, \frac{N_{\text{M}}}{N_{\text{F}}} \}$.
The iterative procedure for BTALS-II is mostly analogous to that summarized for BTALS-I in  Algorithm \ref{alg:BTALS}: we first solve for $\ma{A}$, then we use the solution to solve for $\ma{B}$, and we iterate until the maximum number of iterations or a stop criterion threshold is reached. The only difference with respect to Algorithm~\ref{alg:BTALS} is that we do not force any block of $\hat{\mathbf{B}}^i$ to be an identity matrix.
In fact, BTALS-II is analogous to the BTALS algorithm proposed in \cite{bdris_ce_tensor}, except that the meaning of $\mathring{\ten{H}}$ and ${\ten{T}}$ is different from  that of the corresponding variables in \cite{bdris_ce_tensor}. 

\textit{RBF:}
Our second method is an iterative, reciprocity-aware bilinear factorization (RBF) that is summarized in Algorithm~\ref{alg:RBF}. Indeed, with known $\mathbf{\Omega}_k$, the \textit{Type 2} MC-aware CE problem can be viewed as factorizing the tensor slices $\mathring{\mathbf{H}}_k = \mathbf{A}\,\mathbf{\Omega}_k\,\mathbf{B}$. 
To exploit the known matrices $\mathbf{\Omega}_k$, we work with the mode-3 unfolding in \eqref{eq:type_2_unf_3}, because it separates the dependence on the training configurations from the unknown bilinear factor $(\mathbf{B}^{\top}\otimes\mathbf{A})$. However, $\nmode{\ten{T}}{3}$ cannot be directly pseudo-inverted to recover $(\mathbf{B}^{\top}\otimes\mathbf{A})$ because $\text{rank}(\nmode{\ten{T}}{3}) \leq \text{min}(N_{\text{M}}(N_{\text{M}} +1)/2,\,K)$. Indeed, $\nmode{\ten{T}}{3}= [\mathrm{vec}(\mathbf{\Omega}_1),\ldots,\mathrm{vec}(\mathbf{\Omega}_K)]^{\top}$ contains redundant columns due to $\mathbf{\Omega}_k=\mathbf{\Omega}_k^\top$, which is imposed by the DMA's reciprocity. 
This redundancy can be removed by introducing a duplication matrix. Specifically, $\nmode{\ten{T}}{3}$ can be factorized as $\nmode{\ten{T}}{3} = (\ma{W}\,\bar{\ma{S}})^\top$, where $\ma{W}$ contains $N_{\text{M}}$ columns with a single nonzero entry equal to one, corresponding to the diagonal elements of $\ma{\Omega}_k$, whereas the remaining $\frac{N_{\text{M}}(N_{\text{M}}+1)}{2}-N_{\text{M}}$ columns contain two nonzero entries equal to one, which replicate the symmetric off-diagonal elements. The matrix $\bar{\ma{S}} \in \bb{C}^{\frac{N_{\text{M}}(N_{\text{M}}+1)}{2} \times K}$ collects the independent parameters of the symmetric matrices $\ma{\Omega}_k$, i.e., $\bar{\ma{S}} =
\big[ \text{vech}(\ma{\Omega}_1),\ldots,\text{vech}(\ma{\Omega}_K) \big]$.

Applying a transposition to \eqref{eq:type_2_unf_3} and substituting $\nmode{\ten{T}}{3}^\top = \ma{W}\,\bar{\ma{S}}$ yields
\begin{equation}
\nmode{\mathring{\ten{H}}}{3}^\top =
(\ma{B}^\top\otimes \ma{A})\ma{W}\,\bar{\ma{S}},
\end{equation}
where $\bar{\ma{S}}$ is known and generally has full row rank when $K \geq \frac{N_{\text{M}}(N_{\text{M}}+1)}{2}$. 
After filtering out $\bar{\ma{S}}$, we apply a permutation matrix $\ma{P} \in \bb{R}^{\frac{N_{\text{M}}(N_{\text{M}}+1)}{2} \times \frac{N_{\text{M}}(N_{\text{M}}+1)}{2}}$ so that the diagonal components appear in the first $N_{\text{M}}$ columns:
\begin{align}
   \label{eq:rbf_zf_step} \bar{\ma{H}} & \triangleq \nmode{\mathring{\ten{H}}}{3}^\top\bar{\ma{S}}^+\ma{P} \approx (\ma{B}^\top\otimes \ma{A})\ma{W}\ma{P} \\
  \notag  &= [\ma{b}_1 \otimes \ma{a}_1,\,\, \ldots,\,\,\ma{b}_{N_{\text{M}}} \otimes \ma{a}_{N_{\text{M}}},\,\, \ma{U}],
\end{align}
where $\ma{U}$ contains the combination of the off-diagonal components, i.e., the columns of $\ma{U}$ are  $\text{vec}\!\left(\ma{a}_i\ma{b}_j^\top + \ma{a}_j\ma{b}_i^\top\right)$, for $\{i,j\} = 1,\ldots  N_\mathrm{M}$, with $i<j$. Note that each of the first $N_{\text{M}}$ columns of $\bar{\ma{H}}$ is the vectorization of a rank-one matrix, whereas each of the remaining columns is the vectorization of a matrix of rank at most two. 

Next, let $\ma{M}_{ii}\in\bb{C}^{N_{\text{U}}\times N_{\text{F}}}$ denote the matrix obtained by reshaping the $i$th column of $\bar{\ma{H}}$, for $i=1,\dots,N_{\text{M}}$. Likewise, let $\ma{M}_{ij}\in\bb{C}^{N_{\text{U}}\times N_{\text{F}}}$ denote the matrix obtained by reshaping the column of $\bar{\ma{H}}$ associated with the pair $(i,j)$, for $1\le j<i\le N_{\text{M}}$.
Now, we can formulate the following problem:
\begin{equation}\label{eq:problem_sabf}
\min_{\ma{A},\ma{B}} \left(
\sum_{i=1}^{N_\mathrm{M}}\mu_{ii}
+
\sum_{1\le j<i\le N_\mathrm{M}}
\nu_{ij} \right) ,
\end{equation}
where
\begin{equation}
  \mu_{ii}=\left\|\ma{M}_{ii}-\ma{a}_i\ma{b}_i^\top\right\|_F^2, \quad   \nu_{ij}=\left\|\ma{M}_{ij}-\ma{a}_j\ma{b}_i^\top-\ma{a}_i\ma{b}_j^\top\right\|_{\text{F}}^2.
\end{equation}

Since the objective in (\ref{eq:problem_sabf}) is bilinear in $(\ma{A},\ma{B})$, RBF adopts an alternating minimization strategy. Note that the problem in \eqref{eq:problem_sabf} combines rank-one and rank-two factorizations. Thus, an initial estimate is obtained by considering the first $N_{\text{M}}$ columns of $\bar{\ma{H}}$. By reshaping each of them into a matrix, we obtain blocks of the form $\ma{M}_{ii} \approx \ma{a}_i \ma{b}_i^\top$. This initialization step can be performed using any rank-one factorization method, such as the singular-value decomposition or the power method. However, using only the diagonal terms leads to $N_{\text{M}}$ independent scale ambiguities, i.e., one per column. The off-diagonal blocks resolve these ambiguities by coupling different columns, ultimately leaving only a single global scale factor.

To avoid repeating identical derivations, we now introduce the generic matrices $\ma{X}=[\ma{x}_1,\dots,\ma{x}_{N_{\text{M}}}]
 \in \bb{C}^{d_\mathrm{X}\times N_{\text{M}}} $ and $\ma{Y}=[\ma{y}_1,\dots,\ma{y}_{N_{\text{M}}}] \in \bb{C}^{d_\mathrm{Y}\times N_{\text{M}}}$, where $(\ma{X},\ma{Y})\in\{(\ma{A},\ma{B}^\top),(\ma{B}^\top,\ma{A})\}$ and therefore $ (d_\mathrm{X},d_\mathrm{Y}) \in\{(N_{\text{U}},N_{\text{F}}),(N_{\text{F}},N_{\text{U}})\}$. Then, given  an initialization for $\ma{X}$, the update of $\ma{Y}$ is obtained by solving the normal equation
\begin{equation}
\ma{G}_\mathrm{X} \ma{Y}^{\top} = \ma{R}_\mathrm{Y}^{\top} ,
\label{eq29}
\end{equation}
where the $(i,j)$th entry of $\ma{G}_\mathrm{X}\in\bb{C}^{N_{\text{M}}\times N_{\text{M}}}$ is given by
\begin{equation}
\label{eq:GX_def}
(\ma{G}_\mathrm{X})_{ij} =
\begin{cases}
\sum_{\ell=1}^{N_{\text{M}}} \|\ma{x}_\ell\|_2^2, & i=j,\\
\ma{x}_j^H\ma{x}_i, & i\neq j,
\end{cases}
\qquad i,j=1,\ldots,N_{\text{M}},
\end{equation}
and the $i$th row of $\ma{R}_\mathrm{Y}^\top = [\ma{r}_1^\top,\ldots,\ma{r}_{N_{\text{M}}}^\top]^\top \in \bb{C}^{N_{\text{M}}\times d_\mathrm{Y}}$ is given, for $i=1,\ldots,N_{\text{M}}$, by
\begin{align}
\label{eq:ri_def}
\ma{r}_i^\top =
\ma{x}_i^H \ma{M}_{ii}
+
\sum_{j<i}\ma{x}_j^H \ma{M}_{ij}
+
\sum_{k>i}\ma{x}_k^H \ma{M}_{ki}.
\end{align}
Solving \eqref{eq29} for $\ma{Y}$ yields the update
\begin{equation}
\label{eq:Y_update}
\ma{Y} = \ma{R}_\mathrm{Y}(\ma{G}_\mathrm{X}^{-1})^\top \in \bb{C}^{d_\mathrm{Y}\times N_{\text{M}}}.
\end{equation}
By alternating between updating $\hat{\ma{B}}^\top$ with $\hat{\ma{A}}$ fixed and updating $\hat{\ma{A}}$ with $\hat{\ma{B}}^\top$ fixed, until a maximum number of iterations or a convergence threshold is reached, we ensure that the columns of the estimated matrix are unique up to a single scaling ambiguity, i.e., $\ma{A} = \alpha \hat{\ma{A}}$ and $\ma{B} = \alpha^{-1}\hat{\ma{B}}$, where $\alpha\in\mathbb{C}$ is a nonzero scalar.
This single scaling ambiguity is the same reciprocal scalar ambiguity as in the BTALS case discussed above: for any nonzero scalar $\alpha\in\mathbb{C}$, replacing $\ma{A}$ by $\alpha\ma{A}$ and $\ma{B}$ by $\alpha^{-1}\ma{B}$ leaves $\ma{A}\, \ma{\Omega}_k \,\ma{B}$ unchanged for every $k$. Hence, this ambiguity is operationally irrelevant because it does not affect the reconstructed equivalent channel. If MC vanishes, the off-diagonal coupling terms disappear, the columns decouple, and additional per-column scaling ambiguities remain---again without impacting the prediction of the end-to-end channel.

 \begin{algorithm}[!t]
\caption{Reciprocity-Aware Bilinear Factorization (RBF)}
\label{alg:RBF}
\begin{algorithmic}[1]
\Require $\{\ma{M}_{ii}\}_{i=1}^{N_{\text{M}}}$,  $\{\ma{M}_{ij}\}_{1\le j<i\le N_{\text{M}}}$, max. number of iterations $L_{\max}$.

\State \textbf{Initialization:}
\For{$i=1,\dots,N_{\text{M}}$}
    \State Compute a rank-one factorization of $\ma{M}_{ii}\approx \ma{a}_i\ma{b}_i^\top$.
\EndFor
\State Form $\hat{\ma{A}}=[\hat{\ma{a}}_1,\dots,\hat{\ma{a}}_{N_{\text{M}}}]$, and  $\hat{\ma{B}}^{\text{T}}=[\hat{\ma{b}}_1,\dots,\hat{\ma{b}}_{N_{\text{M}}}]$.
\State \textbf{Main Loop:}
\For{$\ell=1,\dots,L_{\max}$}

    \For{$(\ma{X},\ma{Y})\in\{(\hat{\ma{A}},\hat{\ma{B}}^{\top}),(\hat{\ma{B}}^{\top},\hat{\ma{A}})\}$}
\State Build $\ma{G}_\mathrm{X}$ using \eqref{eq:GX_def}.
\State Form $\ma{R}_\mathrm{Y}^\top$ from $\{\ma{r}_i^\top\}_{i=1}^{N_{\text{M}}}$, with $\ma{r}_i^\top$  given by \eqref{eq:ri_def}.
\State Update $\ma{Y}$ using \eqref{eq:Y_update}.      

    \EndFor
\EndFor
\State \Return $\hat{\ma{A}}$ and $\hat{\ma{B}}^{\top}$
\end{algorithmic}
\end{algorithm}

\subsection{Type 3 MC-Aware CE Problem}\label{subsec_Type3}

\textit{BTALS-III:}\footnote{Because this algorithm can be viewed as a specialized variant of BTALS-I, we coin it BTALS-III for notational convenience. Yet, to be clear, the alternating-LS feature of BTALS-I is not present in BTALS-III.} With known $\mathbf{B}$, we can estimate $\mathbf{H}_0$ and $\mathbf{A}$ using a zero-forcing approach. 
We leverage the same tensor structure as in Sec.~\ref{subsec_Type1}, where $\ma{H}_0$ and $\ma{A}$  are stacked in $\ma{C}$ (see \eqref{eq:Hk_factorized_slice_tucker_compact}). Specifically, we use the $1$-mode unfolding in \eqref{eq:tenH_1} to compute an estimate of $\ma{C} = \left[\mathbf{H}_0,\, \ma{A}\right]$:
\begin{equation}
    \label{eq:type3_A_est_H0_est} \hat{\ma{C}} = \nmode{\ten{H}}{1}\big[\nmode{\bar{\ten{T}}}{1}(\ma{I}_K \otimes \ma{D}^{\top})^\top\big]^+.
\end{equation}
A necessary dimensional condition for unique zero-forcing recovery is $K\geq 1 + \frac{N_{\text{M}}}{N_{\text{F}}}$. Given  $\hat{\ma{C}}$, we identify $\hat{\ma{H}}_0$ and $\hat{\ma{A}}$ based on $\hat{\ma{C}} = [\hat{\ma{H}}_0,\hat{\ma{A}}]$.

\subsection{Type 4 MC-Aware CE Problem}
\label{subsec_Type4}

\textit{BTALS-IV:}\footnote{Because this algorithm can be viewed as a specialized variant of BTALS-I, we coin it BTALS-IV for notational convenience. Yet, to be clear, the alternating-LS feature of BTALS-I is not present in BTALS-IV.} With known $\mathbf{H}_0$ and $\mathbf{B}$, we can estimate $\mathbf{A}$ using a zero-forcing approach. This time, we leverage the same tensor structure as in Sec.~\ref{subsec_Type2}. Specifically, we use the 1-mode unfolding in \eqref{eq:type_2_unf_1} to compute an estimate of $\mathbf{A}$:
\begin{equation}
    \label{eq:type4_A_est} \hat{\ma{A}} = \nmode{\mathring{\ten{H}}}{1}\big[\nmode{\ten{T}}{1}(\ma{I}_K \otimes \ma{B}^{\top})^\top\big]^+.
\end{equation}
A necessary dimensional condition for the uniqueness of the pseudo-inverse in \eqref{eq:type4_A_est} is $K\geq \frac{N_{\text{M}}}{N_{\text{F}}}$.

\section{Experimental Validation}
\label{sec_ExpVal}

\subsection{DMA Prototype}
\label{subsec_prototype}

Our DMA is an ultrathin device that couples eight feeds (of which we use only seven, as explained below) to 96 radiating meta-elements. The coupling structure is a quasi-2D irregularly shaped cavity, similar to the one described in~\cite{sleasman2020implementation}. The motivation for choosing this coupling structure is that it induces strong MC between the meta-elements, such that the DMA can reap the increased wave-domain flexibility enabled by strong inter-element MC that was unveiled in~\cite{prod2024mutual,MCbenefitsDMA}. In contrast to microstrip-based coupling structures that deliberately mitigate MC with large inter-element spacing~\cite{boyarsky2021electronically}, accounting for MC is pivotal for our DMA prototype, as we show below.

\begin{figure}
    \centering
    \includegraphics[width=\columnwidth]{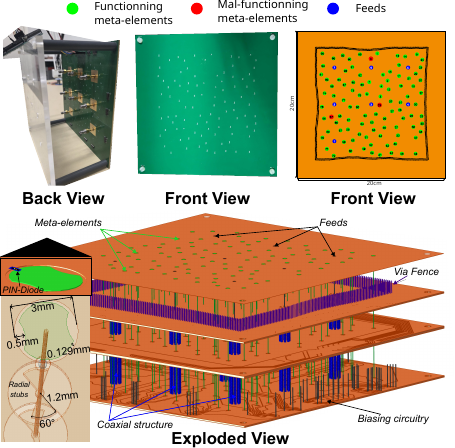}
    \caption{(a) Photographic images of front view and back view of our DMA with eight feeds (of which we use only seven, as explained in Sec.~\ref{subsec_prototype}) and 96 1-bit-programmable meta-elements. (b) Front-view overview of the location of the feeds, meta-elements, and via fence. (c) Exploded view showing the bias circuitry in the interior of the four-layer PCB.}
    \label{Fig2}
\end{figure}

More specifically, our DMA is a four-layer PCB, as seen in Fig.~\ref{Fig2}c. The approximately  $15\ \mathrm{cm} \times 15\ \mathrm{cm}$ large cavity is formed in the 1.52~mm thick top layer filled with low-loss dielectric Rogers 4003 substrate ($ \epsilon_{r} = 3.55$, $\mathrm{tan}\delta = 0.019$) using a fence comprising 808 vias. The cavity can be excited by eight coaxial feeds from the back of the PCB: each RF connector's pin is connected to a via that passes through the PCB to a circular patch on the front. The outer conductor of the connector is connected to the RF ground. The circular patch is separated from the ground by an annular gap enabling matching of the feed and the cavity. The 96 meta-elements are pseudo-randomly placed on the top layer, with the condition to be at least 1.5~cm apart from one another. All meta-elements have the same orientation. The meta-elements consist of a rounded complementary electric-LC resonator that can be tuned using a PIN diode as described in~\cite{yoo2016efficient}.

Each meta-element can be individually configured to two possible states by applying a binary bias voltage (0~V or 5~V) to its PIN diode using a via that traverses the PCB and is connected to the control circuit. The control circuit is implemented in the two bottom layers of the PCB, consisting of twelve eight-bit shift-registers and four buffers. An Arduino Mega reconfigures the shift registers when it receives a 96-element binary control vector $\mathbf{v}$ from Python.

The description of the PIN diodes as tunable lumped elements is justified because of their small size (0.457~mm) relative to the free-space wavelength at 18~GHz (16.67~mm) and the wavelength in the Rogers 4003 dielectric (8.85~mm); see Sec.~IV.C in~\cite{tapie2025experimental} for related experimental sanity checks. 
Throughout our experiments, we only use seven of the DMA's eight feeds. The reason is simply that we use an eight-port VNA with one port connected to a probe, as explained below, such that we can connect only seven VNA ports to the DMA feeds.
The remaining eighth feed is left open-circuited in all experiments, and thus acts like an additional scatterer in the already chaotic cavity of our DMA.
Altogether, we thus have $N_\mathrm{F}=7$ and $N_\mathrm{M}=96$. Three meta-elements are non-functional due to fabrication issues (missing diodes) and indicated in red rather than green color in Fig.~\ref{Fig2}b; however, we do not handle them differently from functioning meta-elements, evidencing that our algorithms do not require prior knowledge of faulty meta-elements.

\begin{figure*}
    \centering
    \includegraphics[width=\textwidth]{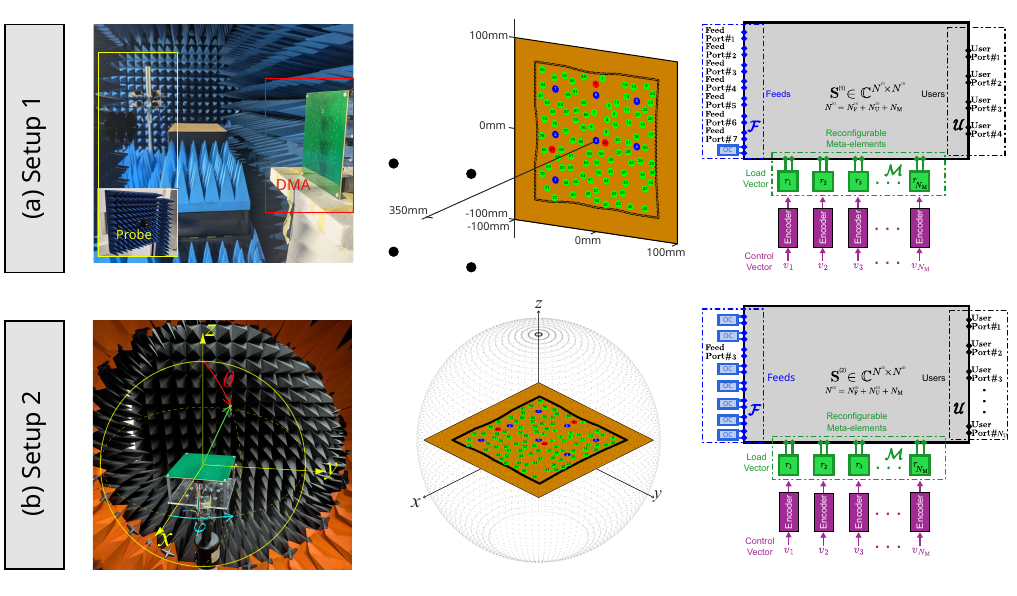}
    \caption{(a) Experimental Setup 1 with $N_\mathrm{F}=7$ and $N_\mathrm{U}=4$. The VAA has four grid points in a plane parallel to the DMA surface. The VAA is realized by moving a probe on a translation stage. (b) Experimental Setup 2 with $N_\mathrm{F}=1$ and $N_\mathrm{U}=10976$. The VAA has 5488 dual-polarization grid points (i.e., 10976 user ports) on a sphere whose center coincides with the DMA's center. The VAA is realized by rotating the DMA within an arc of dual-polarized probes. In both setups, the eighth DMA port is open-circuited throughout all experiments. }
    \label{Fig3}
\end{figure*}

\subsection{Experimental Setups and Measurement Procedure}
\label{subsec_setupANDprocedure}

In our experiments, we sample the field radiated by the DMA at ensembles of discrete points with ``virtual antenna arrays'' (VAAs). A VAA is realized by mechanically displacing one or a few probe(s) relative to the DMA, or vice versa. For single-polarization probes, each VAA grid point represents one user port. For dual-polarization probes, each VAA grid point is represented by two user ports.

\textit{Setup 1:} As shown in Fig.~\ref{Fig3}a, our DMA is placed inside an anechoic chamber, and the VAA is realized based on a 2D translation stage scanning a single-polarization probe through four grid points (a square with side length 10~cm) in a plane parallel to, and 35~cm in front of, the DMA's surface. Thus, we have $N_\mathrm{U}=4$. The probe's polarization is aligned with the dominant polarization of the meta-elements. Seven DMA feed ports and the probe are connected to an eight-port vector network analyzer (VNA). We measure at a single frequency (18~GHz) with VNA settings intended to maximize the measurement accuracy (power:~13~dBm,~IFBW:~500~Hz).

Similar to~\cite{tapie2025experimental}, we first measure the DMA's $7\times 7$ reflection matrix for 1000 random DMA configurations. Then, we measure the end-to-end channel matrix from the DMA's feeds to the VAA grid points for the reference DMA configuration $\mathbf{v}_0$, 96 single-flip DMA configurations (the $i$th single-flip configuration differs from the reference configuration only regarding the state of the $i$th meta-element), and 10000 random configurations. To be clear, our proposed CE methods only directly use a subset of the 10000 measurements of the end-to-end channel matrix with random DMA configurations. We measure that many random DMA configurations to conduct ablation studies. The other types of measurements (all reflection-matrix measurements, and end-to-end channel matrix measurements with reference and single-flip configurations) are only required for the technique described in~\cite{tapie2025experimental} that we run once to subsequently have prior knowledge of some model parameters, as discussed in Sec.~\ref{sec_ProblemStatement}.
Based on the resulting consistent set of proxy parameters $\tilde{\alpha}$, $\tilde{\beta}$, and $\tilde{\mathbf{\Gamma}}$, we compute $\tilde{\mathbf{\Omega}}_k$ in the corresponding ambiguity basis using \eqref{eq:Sk_def} for every measured DMA configuration $\mathbf{v}_k$. The technique in~\cite{tapie2025experimental} enforces reciprocity on $\tilde{\mathbf{\Gamma}}$, such that $\tilde{\mathbf{\Omega}}_k$ is symmetric: $\tilde{\mathbf{\Omega}}_k = \tilde{\mathbf{\Omega}}_k^\top$.

We monitor the stability of our experiment over the course of the measurements by intermittently re-measuring $\mathbf{H}$ for the same random DMA configuration. To quantify the stability, we treat the difference between the measurements of $\mathbf{H}$ under nominally identical conditions as noise and evaluate an effective signal-to-noise ratio (SNR). Over the course of our measurements with Setup 1, this stability metric does not drop below 43.3~dB.

\textit{Setup 2:}
As shown in Fig.~\ref{Fig3}b, our DMA is placed on a rotatable platform such that the DMA's center coincides with the center of an arch of dual-polarization probes inside an anechoic environment. In this commercial StarLab setup, only the central feed (indexed \#3 in Fig.~\ref{Fig2}) is connected to the VNA, while the other feeds are left open-circuited (for which we duly account, as explained in the next paragraph). The StarLab setup sequentially connects the probe ports to one VNA port and controls the rotation of the DMA relative to the probe arch. The acquired data are post-processed by the StarLab software to reconstruct the complex dual-polarization field on a spherical surface of diameter 25~cm centered on the DMA (see Fig.~\ref{Fig3}b for the angle convention). The result is provided on an angular grid of $56\times 98$ directions, which we interpret as a spherical VAA: each grid direction corresponds to a virtual user equipped with a non-invasive dual-polarized probe located on that sphere. Considering the two polarizations, we thus have $N_\mathrm{U}=2\cdot 56\cdot 98=10976$ user ports.
In other words, for each DMA configuration, the end-to-end channel vector contains the complex transfer functions that would be observed by dual-polarized receive probes at those spherical sampling directions.

Our analysis of the data from Setup 2 relies on the estimates of $\tilde{\alpha}$, $\tilde{\beta}$, and $\tilde{\mathbf{\Gamma}}$ obtained with the method from~\cite{tapie2025experimental} based on Setup 1. 
To account for the fact that only feed \#3 is connected in Setup~2 while the other six feed ports (connected in Setup~1) are left open-circuited in Setup~2, we compute an effective MC matrix by eliminating the open feed ports via a standard MNT port-termination reduction. Specifically, using the proxy scattering-matrix blocks $\tilde{\mathbf{S}}$ identified from Setup~1, let $\mathcal{F}_\mathrm{act}=\{3\}$ denote the active feed port and $\mathcal{F}_\mathrm{oc}=\mathcal{F}\setminus\mathcal{F}_\mathrm{act}$ the set of open-circuited feed ports, and partition $\tilde{\mathbf{S}}$ accordingly. With open-circuit terminations (reflection coefficient $+1$ w.r.t.\ $Z_0$), the multiple-scattering contribution through $\mathcal{F}_\mathrm{oc}$ can be absorbed into an effective MC matrix
\begin{equation}
\tilde{\mathbf{\Gamma}}_\mathrm{eff}
=\tilde{\mathbf{S}}_{\mathcal{MM}}
+\tilde{\mathbf{S}}_{\mathcal{M}\mathcal{F}_\mathrm{oc}}
\bigl(\mathbf{I}-\tilde{\mathbf{S}}_{\mathcal{F}_\mathrm{oc}\mathcal{F}_\mathrm{oc}}\bigr)^{-1}
\tilde{\mathbf{S}}_{\mathcal{F}_\mathrm{oc}\mathcal{M}},
\end{equation}
which we then use as the known MC matrix in \eqref{eq_MNT} for Setup~2.

Over the course of our measurements with Setup 2, our stability metric does not drop below 28.9~dB. The lower stability of the measurements with Setup 2 (in comparison to Setup 1) might originate from vibrations associated with the motion of the arch of probes in the StarLab setup. 
In addition, we note that the reliance on the values of $\tilde{\alpha}$, $\tilde{\beta}$, and $\tilde{\mathbf{\Gamma}}$ obtained with Setup 1 implies a potential source of mild model-reality mismatch which can arise due to mechanical stress and deformations arising from manual interventions in moving the DMA from Setup 1 to Setup 2, as well as due to temperature and humidity changes over the course of a one-month gap between the measurements on Setup~1 and Setup~2.

\subsection{Experimentally Verifiable CE Accuracy Metrics}
\label{subsec:acc_metrics}

Since the ground-truth parameter values are unknown and not directly observable, we evaluate the CE performance of our algorithms exclusively through \emph{physically measurable} quantities. Concretely, we measure the end-to-end channel matrix for a set of $Q$ unseen, randomly chosen DMA configurations and compare the measurements to the corresponding model-based predictions. We evaluate two complementary metrics.

We define the normalized mean-squared error (NMSE) as 
\begin{equation}
\label{eq:nmse_global}
\mathrm{NMSE}
\triangleq
\frac{\sum_{q=1}^{Q}\fronorm{\mathbf{H}^{(q)}_{\mathrm{meas}}-\mathbf{H}^{(q)}_{\mathrm{pred}}}^{2}}
{\sum_{q=1}^{Q}\fronorm{\mathbf{H}^{(q)}_{\mathrm{meas}}}^{2}},
\end{equation}
where $\mathbf{H}^{(q)}_{\mathrm{meas}}$ and $\mathbf{H}^{(q)}_{\mathrm{pred}}$ denote, respectively, the measured and predicted end-to-end channel matrix for the $q$th test configuration.
While the NMSE is a standard metric, it is sensitive to the magnitude of the constant term in the system model in (\ref{eq_MNT}). We expect $\mathbf{H}_0$ to be small for our DMA architecture, but the parameter ambiguities nonetheless allow $\tilde{\mathbf{H}}_0$ in principle to be substantial. Therefore, we also evaluate a second accuracy metric that is insensitive to any configuration-independent offset across the test index $q$. 
Following~\cite{rabault2023tacit,tapie2025experimental}, we define
\begin{equation}
\label{eq:zeta_def}
\zeta
\triangleq
\left\langle
\frac{\operatorname{SD}_q\!\Big(H^{(q)}_{\mathrm{meas},ij}\Big)}
{\operatorname{SD}_q\!\Big(H^{(q)}_{\mathrm{meas},ij}-H^{(q)}_{\mathrm{pred},ij}\Big)}
\right\rangle_{i,j},
\end{equation}
where $H^{(q)}_{\mathrm{meas},ij}$ and $H^{(q)}_{\mathrm{pred},ij}$ denote the $(i,j)$th entries of $\mathbf{H}^{(q)}_{\mathrm{meas}}$ and $\mathbf{H}^{(q)}_{\mathrm{pred}}$, respectively, $\operatorname{SD}_q(\cdot)$ is the standard deviation over the test index $q$, and $\langle\cdot\rangle_{i,j}$ denotes averaging over all matrix indices.

\subsection{Experimental CE Results}

\begin{figure*}
    \centering
    \includegraphics[width=\textwidth]{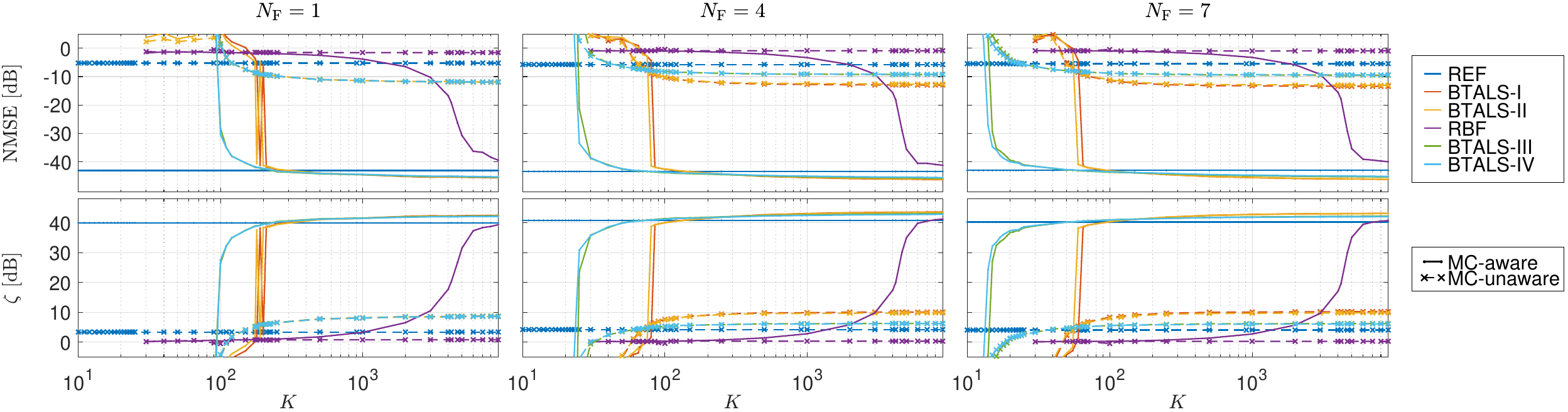}
    \caption{Based on measurements on Setup 1 shown in Fig.~\ref{Fig3}a, two CE accuracy metrics (first row: NMSE defined in \eqref{eq:nmse_global}; second row: $\zeta$ defined in \eqref{eq:zeta_def}) are plotted as a function of $N_\mathrm{F}$ (different columns) and $K$ (horizontal axis), for the five different algorithms described in Sec.~\ref{sec_CEalgorithms} (BTALS-I, BTALS-II, RBF, BTALS-III, BTALS-IV), with and without MC awareness. The CE accuracy metrics achieved with the algorithm from~\cite{tapie2025experimental} are plotted as reference (REF).}
    \label{Fig4}
\end{figure*}

\begin{table}[b]
\caption{Comparison of minimum required training overhead $K_\mathrm{min}^\mathrm{EXP}$ based on Fig.~\ref{Fig4}.}
\label{tab:trainingoverhead}
\centering
\renewcommand{\arraystretch}{1.5}
\setlength{\tabcolsep}{4pt}
\begin{tabular}{|c||*{5}{>{\centering\arraybackslash}p{0.145\columnwidth}|}}
\hline
$N_\mathrm{F}$ & BTALS-I & BTALS-II & RBF & BTALS-III & BTALS-IV \\
\hline
1 & 289 & 314 & $>9000$ & 229 & 235 \\
4 & 134 & 137 & 7500 & 72 & 73 \\
7 & 100 & 100 & 7000 & 50 & 50 \\
\hline
\end{tabular}
\end{table}

We now systematically examine in Fig.~\ref{Fig4} the performance and training overhead of the five algorithms presented in Sec.~\ref{sec_CEalgorithms} based on our experimental measurements on Setup 1 (shown in  Fig.~\ref{Fig3}a). To that end, we evaluate NMSE and $\zeta$ (defined in Sec.~\ref{subsec:acc_metrics}) as a function of $K$ and $N_\mathrm{F}$. For reference, we also evaluate these metrics using the proxy parameter set obtained via the procedure described in~\cite{tapie2025experimental} (denoted by REF). In addition, we benchmark against an MC-unaware scenario for which we impose that $\tilde{\mathbf{\Gamma}}$ vanishes. In addition to the achieved accuracy (quantified in terms of NMSE and $\zeta$), we also assess the minimum required training overhead. For simplicity, we quantify the minimum required training overhead as the minimum number of measurements $K_\mathrm{min}^\mathrm{EXP}$ for which the value of $\zeta$ reaches the corresponding REF value. A summary of the experimentally determined values of $K_\mathrm{min}^\mathrm{EXP}$ is provided in Table~\ref{tab:trainingoverhead}. The noise level is the experimentally given one on the order of 43.4~dB, as discussed in Sec.~\ref{subsec_setupANDprocedure}.

The accuracy achieved by the REF proxy parameters is around $\zeta=40$~dB (39.9~dB for $N_\mathrm{F}=1$, 40.8~dB for $N_\mathrm{F}=4$, 40.2~dB for $N_\mathrm{F}=7$). If MC is neglected, these values drop to around $\zeta=4$~dB, highlighting the importance of accounting for MC. With the largest considered value of $K=9000$, all five considered algorithms achieve accuracies that exceed REF for $N_\mathrm{F}=7$. RBF marginally exceeds REF by 0.5~dB, BTALS-IV and BTALS-III exceed REF by 1.8~dB and 1.9~dB, respectively, and BTALS-II and BTALS-I both exceed REF by 2.9~dB. BTALS-I/II reach accuracies of $\zeta=43.1$~dB, which is extremely close to the upper bound of 43.4~dB imposed by the noise level in our experiment.
The ability of the five algorithms to exceed REF originates from their superior noise rejection enabled by the tensor approach. It is interesting to note the hierarchy among the four BTALS algorithms in terms of how much they exceed REF for $K=9000$. The fewer parameters are assumed known, i.e., the more parameters BTALS estimates, the more the REF performance can be exceeded. Similar observations can be made for $N_\mathrm{F}=4$. For $N_\mathrm{F}=1$, RBF has not yet fully reached the REF accuracy with $K=9000$ (its accuracy is 0.6~dB below REF), and the accuracies of the four BTALS algorithms exceed the REF accuracy by 2.2~dB to 2.6~dB (with the same hierarchy discussed before).

In terms of the minimum required training overhead, the five algorithms clearly cluster into three groups. BTALS-III and BTALS-IV have the lowest overheads, BTALS-I and BTALS-II have medium overheads, and RBF has a very large overhead. It makes sense that the overhead for BTALS-III and BTALS-IV is much lower since they estimate many fewer parameters. As documented in Table~\ref{tab:trainingoverhead}, the overhead for BTALS-I/II is twice the overhead for BTALS-III/IV, while the overhead for RBF is more than an order of magnitude larger than for BTALS-I/II. For lower values of $N_\mathrm{F}$, the minimum required overhead increases. For BTALS-I, $K_\mathrm{min}^\mathrm{EXP}$ increases from 100 for $N_\mathrm{F}=7$, via 134 for $N_\mathrm{F}=4$, to 289 for $N_\mathrm{F}=1$. 

The inferior performance of the RBF algorithm can be explained by the zero-forcing in its first step in \eqref{eq:rbf_zf_step}. A necessary dimensional condition is $K\geq \frac{N_{\text{M}}(N_{\text{M}} +1)}{2}$, which explains RBF's high values of $K_\mathrm{min}^\mathrm{EXP}$. Moreover, since $\bar{\ma{S}}$ is not orthogonal, the initial zero-forcing step can amplify and color the noise, thereby degrading performance. Nonetheless, RBF has one notable strength: RBF computes rank-one and rank-two factorizations in a bilinear fashion, which significantly reduces computational complexity compared to the BTALS algorithm (Algorithm~\ref{alg:BTALS}) that computes two major matrix inversions per iteration.

We also evaluated the performance of our algorithms under more severe noise conditions by adding synthetic white Gaussian noise to our experimental measurements. These studies yielded qualitatively very similar trends to those seen in Fig.~\ref{Fig4} except that the lower [upper] bound on the achievable NMSE [$\zeta$] was increased [decreased] by the addition of the synthetic noise (which also impacts $\mathbf{H}_\mathrm{meas}^{(q)}$ in \eqref{eq:nmse_global} and \eqref{eq:zeta_def}).

\begin{figure}
    \centering
    \includegraphics[width=0.65\columnwidth]{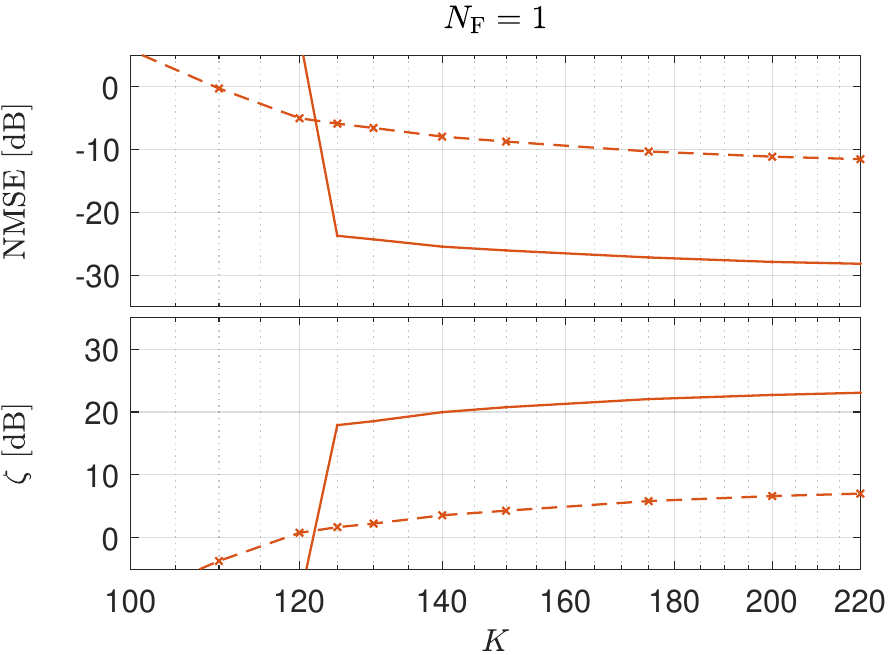}
    \caption{Based on measurements on Setup 2 shown in Fig.~\ref{Fig3}b, two CE accuracy metrics (first row: NMSE defined in \eqref{eq:nmse_global}; second row: $\zeta$ defined in \eqref{eq:zeta_def}) are plotted as a function of $K$ (horizontal axis) for the BTALS-I algorithm, with and without MC awareness.}
    \label{Fig5}
\end{figure}

Finally, we also test the BTALS-I algorithm on experimental data measured on Setup 2 (shown in Fig.~\ref{Fig3}b). The achieved accuracy as a function of $K$ is plotted in Fig.~\ref{Fig5}, together with the MC-unaware benchmark. We observe a very sharp increase in accuracy as we increase $K$ from 120 to 125, and thereafter, the accuracy improves only marginally. When $K=220$, we achieve $\zeta=23.1$~dB. This is below the noise-imposed limit of 28.9~dB but plausible given the manual interventions and one-month gap between the experiments on Setup 1 and Setup 2, since the estimates of $\tilde{\alpha}$, $\tilde{\beta}$, and $\tilde{\mathbf{\Gamma}}$ are based on measurements on Setup 1.

\begin{figure*}
    \centering
    \includegraphics[width=\textwidth]{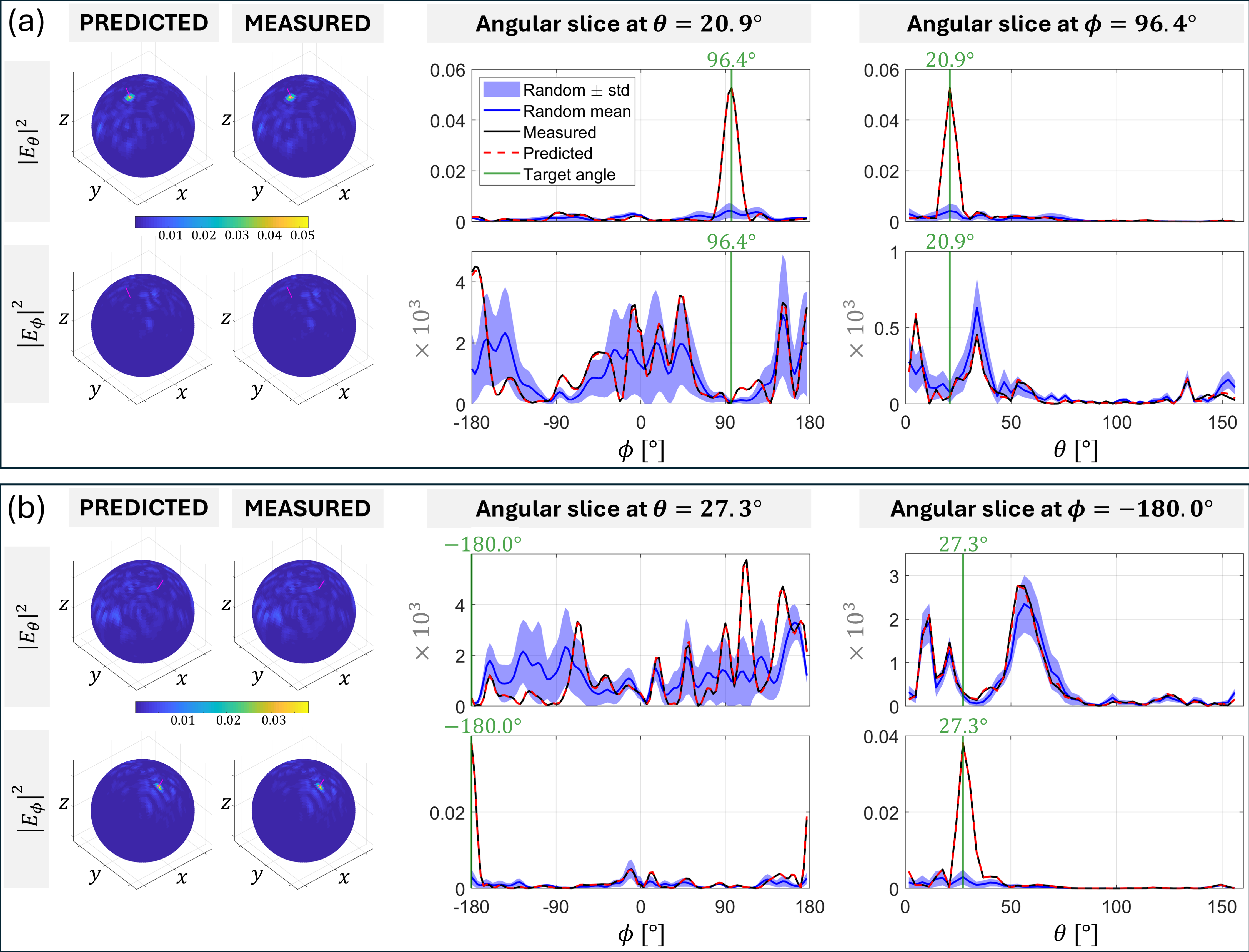}
    \caption{Two selected examples of experimental validations of optimized DMA configurations for beam-steering in a desired direction. On the left, the predicted (left) and measured (right) radiation patterns with the optimized DMA configuration are shown for the $\theta$ (top) and $\phi$ (bottom) field components; a thin magenta line indicates the target direction. On the right, the predicted and measured radiation patterns of the optimized DMA configuration are compared in terms of two angular slices against the mean and standard deviation of 100 random DMA configurations; a green line indicates the target direction. Key metrics are summarized in Table~\ref{tab:target_direction_summary}.  }
    \label{Fig_Opti}
\end{figure*}

\begin{table*}[t]
\centering
\caption{Summary of key metrics from Fig.~\ref{Fig_Opti}.}
\label{tab:target_direction_summary}
\renewcommand{\arraystretch}{1.15}
\setlength{\tabcolsep}{5pt}
\begin{tabular}{c c c c c c c c}
\hline
Fig.~\ref{Fig_Opti} &
{Target pol.} &
{Target dir. } $\theta$  & 
{Target dir.} $\phi$ &
{RAND} &
{OPT MEAS} &
{OPT PRED} &
{Channel gain enhancement} \\
\hline
(a) & $\lvert E_{\theta}\rvert^2$ & $20.9^\circ$ & $96.4^\circ$ & $(4.21\pm2.93)\times10^{-3}$ & $5.26\times10^{-2}$  & $5.23\times10^{-2}$  & 12.5  \\
(b) & $\lvert E_{\phi}\rvert^2$   & $27.3^\circ$ & $-180.0^\circ$  & $(2.99\pm1.95)\times10^{-3}$ & $3.81\times10^{-2}$  & $3.81\times10^{-2}$ & 12.7  \\
\hline
\end{tabular}
\end{table*}

\subsection{Experimental SISO Performance Evaluation}

Although the focus of this paper is on CE, we briefly describe two selected examples of experimental performance evaluations based on the estimated model for Setup 2 in this subsection. Thereby, we directly evidence the usability of the estimated model for model-based optimization. Moreover, we thereby demonstrate that our estimated model accurately predicts the end-to-end channels not only for random DMA configurations but also for optimized DMA configurations.

For simplicity, we consider the SISO channel gain for a target direction and polarization as the key performance indicator (KPI). In the SISO case, the channel gain maps monotonically to system-level KPIs like bit-error rate and capacity. Optimizing the DMA configuration for SISO channel gain maximization is a high-dimensional, non-convex, discrete optimization problem. Various MC-aware discrete optimization algorithms for this problem were studied in the context of RISs in~\cite{hammami2025statistical}. Here, we leverage a standard genetic algorithm for simplicity to perform the required discrete optimization based on our estimated model, using the negative of the target SISO channel gain as the fitness function. We use default parameters (population size: 200, maximum number of generations: $100 N_\mathrm{M}$) and stop early if there is no improvement for the last 50 generations or if the improvement drops below a threshold of $10^{-6}$. 

For the two selected targets, we display the comparison between predicted and measured end-to-end channels in Fig.~\ref{Fig_Opti} and summarize key metrics in Table~\ref{tab:target_direction_summary}. In both cases, the agreement between our model's prediction and the measurement is flawless. For benchmarking, we also show the mean and standard deviation of the channel gain across 100 randomly selected DMA configurations. The channel gain enhancement (measured as the channel gain with the optimized configuration relative to the mean channel gain with random configurations) exceeds 12 in both examples.

\section{Discussion}
\label{sec_Discussion}

Although we focused on conventional DMAs for our experimental validation in this paper, the proposed framework applies directly to any wireless system with wave-domain programmability arising from tunable lumped elements with known MC. Therefore, our method can be directly applied not only to RISs but also to real-world implementations of BD-RISs~\cite{del2025physics} and BD-DMAs~\cite{prod2025beyond}. For BD-RISs and BD-DMAs, the number of tunable lumped elements in the ``beyond-diagonal'' load network differs from the number of meta-elements, and $N_\mathrm{M}$ represents the former; in addition, since the channel matrices represent the propagation between antenna ports and tunable lumped elements, they also account for static components within the ``beyond-diagonal'' load network. Furthermore, our method also directly applies to stacked and flexible embodiments of RISs and DMAs.

\section{Conclusion}
\label{sec_Conclusion}

To summarize, we developed and experimentally validated tensor-decomposition-based CE techniques for DMAs with known MC based on an electromagnetically consistent multiport-network model. We formulated the MC-aware CE problem as a structured tensor factorization problem and proposed a generalized BTALS framework, specialized variants thereof for different levels of prior parameter knowledge, as well as a reciprocity-aware bilinear factorization method. Based on experimental measurements with an 18~GHz DMA prototype in two distinct setups, we showed that accounting for MC is essential and that the proposed tensor-based methods can outperform the reference technique from~\cite{tapie2025experimental} thanks to their superior noise rejection. We systematically examined the minimum required training overhead and achieved accuracy. Finally, we demonstrated that the estimated model supports successful model-based DMA optimization, including the prediction and experimental validation of optimized radiation patterns. These results highlight the practical relevance of MC-aware tensor methods for programmable wave-domain systems such as DMAs, RISs, and related architectures based on tunable lumped elements.

\section*{Acknowledgment}
The authors acknowledge IETR's QOSC test facility (which is part of the CNRS RF-Net network).

\bibliographystyle{IEEEtran}

\providecommand{\noopsort}[1]{}\providecommand{\singleletter}[1]{#1}%

\end{document}